# Transcritical Mixing and Auto-Ignition of n-dodecane Liquid Fuel using a Diffused Interface Method

Luis Bravo[1], Peter C. Ma[2], Matthias Ihme[3]

[1]*Vehicle Technology Directorate, US Army Research Laboratory, Aberdeen Proving Ground, MD 21005*

[2-3]*Mechanical Engineering Department, Stanford University, Stanford, CA 94305*

**High-fidelity simulations of transcritical spray mixing and auto ignition in a combustion chamber are performed at high pressure and temperature conditions using a recently developed finite rate LES solver. The simulation framework is based on a diffused-interface (DI) method that solves the compressible multi-species conservation equations along with the Peng Robinson state equation and real-fluid transport properties. A finite volume approach with entropy stable scheme is employed to accurate simulate the non-linear real fluid flow. LES analysis is performed for non-reacting and reacting spray conditions targeting the ECN Spray A configuration at chamber conditions with a pressure of 60 bar and temperatures between 800 K and 1200 K to investigate effects of the real-fluid environment and low-temperature chemistry. Comparisons with measurements in terms of global spray parameters and mixture fraction distributions demonstrates the accuracy in modeling the turbulent mixing behavior. Good overall agreement of the auto-ignition process is obtained from simulation results at different ambient temperature conditions and the formation of intermediate species is captured by the simulations, indicating that the presented numerical framework adequately reproduces the corresponding low-and-high-temperature ignition processes under high-pressure conditions that are relevant to realistic diesel fuel injection systems.**

## I.    Introduction

As propulsion systems continue to push towards extreme conditions, the demand for high-performance combustion devices operating at ever-higher pressures has been steadily increasing. This trend concerns traditional applications, such as diesel or gas turbine engines, as well as advanced novel designs such as pressure gain (detonation based) combustion systems. In these environments, reacting flows undergo exceedingly complex and interrelated thermophysical processes starting with compressible fuel injection, atomization, heating, and turbulent mixing leading to combustion and are often subject to pressures levels well above the thermodynamic critical point of fluids. In the context of diesel engines, the compressed fluid is typically injected at sub-critical temperatures into a mixture consisting of reactants and products that undergoes a thermodynamic state transition into the supercritical regime. This process is referred to as a transcritical injection, which is schematically shown in Fig. 1 in comparison with a classical subcritical injection process. Mani et al [1] conducted high-speed long-distance microscopy imaging on n-dodecane sprays at diesel operating conditions and showed that the interfacial behavior of fluids exhibits properties markedly different from classical two-phase breakup including diminishing effects from surface tension and heat of vaporization. Dahms et al [2,3] presented a theoretical framework that explains the conditions under which a multi-component fluid mixture transitions from two-phase breakup to single phase mixing in a manner consistent with experimental observations. The theory is based on coupled real fluid thermodynamics and Linear Gradient Theory that was applied to derive a Knudson number criterion and a new regime diagram [2]. The findings suggest that transition to a supercritical state stems as a result of thermal gradients within the interfacial region broadening of the interface and reduced molecular mean free path. The dynamics of the interfacial forces were also shown to gradually decrease as the interface broadens once it enters the continuum regime. This generalized framework provides a theoretical foundation to better understand and model the fuel injection interfacial transition dynamics in regions above the fluid critical points.

A number of recent works [4, 5, 6, 7, 8] have provided insights into the high-pressure fuel injection dynamics using high fidelity large eddy simulation (LES) approaches and experimental techniques. In non-reacting sprays,

---

Ofelein et al [4,5] presented a LES framework coupled with real-fluid thermodynamics and transport to show that for typical diesel conditions (denoted as Spray-A and Spray-H) the dense liquid jet mixing layer dynamics are dominated by non-ideal real fluid behavior. His findings showed that in the case of typical diesel conditions the mixing path associated with all states during injection never crossed the liquid-vapor regime, suggesting that interfacial mixing layer dynamics are locally supercritical. Using similar numerical techniques, Lacaze el al [5] further showed mixing processes that were significantly modified by the real-fluid thermodynamics. It was suggested that the large density ratio between the supercritical fuel and the ambient gas leads to significant penetration of the jet with enhanced turbulent mixing at the tip and strong entrainment effects. Also reported was the presence of supersonic regions in the mixing layer due to the significant decrease in the speed of sound due to the real-gas thermodynamics, turbulent mixing, and high injection velocity. In reacting sprays, using an LES coupled with a Flamelet Generated Manifold (FGM) combustion model, Wehrfritz et al [6] carried out an investigation and characterized the early flame development with respect to temperature and formaldehyde at near Spray-A conditions. He reported auto-ignition taking place in multiple stages denoted as first, second, and flame development stage, and observed formaldehyde formation prior to ignition at the tip of the fuel-rich gas jet consistent with experiments [7]. More recently, a coupled LES and Tabulated Flamelet Model (TFM) with multiple realizations was presented in [8] to study the structure of the flame dynamics at low-temperature engine conditions in the range between 800K-1100K. Significant differences in flame structure were found at the low temperature conditions (800K) including the formation of formaldehyde occurring in lean regions (phi<1). At typical engine conditions (>900K), formaldehyde formation is observed in the rich region followed by OH and high temperature in the stoichiometric region with higher scalar dissipation rates. These findings have important implications that should be considered for further investigations of auto-ignition physics in fuel injection dynamics.

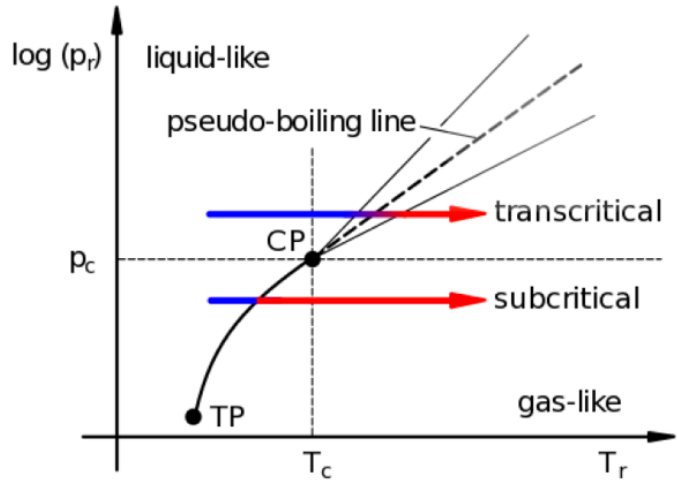

**Figure 1.** Schematic of subcritical and transcritical injection process in a $P_r$-$T_r$ diagram. $C_p$ denotes the critical point and $T_p$ refers to the triple point.

The focus of this work is to investigate the influence of real-fluid thermodynamics on mixing, ignition and combustion process in diesel sprays under compression ignition conditions. A recently developed LES finite rate solver and diffused interface (DI) method [9,10,11] is employed to simulate fuel injection dynamics. The governing equations will be presented followed by the thermodynamic and transport models used for the description of the transcritical turbulent flows. Then, the numerical methods employed in this work will be presented and discussed. The simulation results are compared to the ECN benchmark experimental results including non-reacting and reacting conditions at intermediate to high operating temperature conditions to examine the performance of the thermodynamic and transport models in regions above the fluid critical points.

## II.    Computational Framework

### 2.1 Governing Equations

The governing equations for the diffuse-interface method are the Favre-filtered conservation laws for mass, momentum, total energy, and species, taking the following form,

$$\frac{\partial \bar{\rho}}{\partial t} + \nabla \cdot (\bar{\rho}\tilde{\boldsymbol{u}}) = 0 \, , \tag{1a}$$

$$\frac{\partial \bar{\rho}\tilde{\boldsymbol{u}}}{\partial t} + \nabla \cdot (\bar{\rho}\tilde{\boldsymbol{u}}\tilde{\boldsymbol{u}} + \bar{p}\boldsymbol{I}) = \nabla \cdot \bar{\boldsymbol{\tau}}_{v+t} \, , \tag{1b}$$

$$\frac{\partial \bar{\rho}\tilde{E}}{\partial t} + \nabla \cdot \left[\tilde{\boldsymbol{u}}(\bar{\rho}\tilde{E} + \bar{p})\right] = \nabla \cdot (\bar{\boldsymbol{\tau}}_{v+t} \cdot \tilde{\boldsymbol{u}}) - \nabla \cdot \bar{\boldsymbol{q}}_{v+t} \, , \tag{1c}$$

$$\frac{\partial \bar{\rho}\tilde{Y}_k}{\partial t} + \nabla \cdot (\bar{\rho}\tilde{\boldsymbol{u}}\tilde{Y}_k) = -\nabla \cdot \bar{\boldsymbol{J}}_{k,v+t} + \bar{\dot{\omega}}_k \, , \tag{1d}$$

where $\rho$ is the density, $\boldsymbol{u}$ is the velocity vector, $p$ is the pressure, $e_t$ is the specific total energy, and $Y_k$, $\boldsymbol{j}_k$ and $\dot{\omega}_k$ are the mass fraction, diffusion flux, and chemical source term for species $k$, and the species equations are solved for $k = 1$, ...., $N_S$ - 1 where $N_S$ is the number of species.

The viscous stress tensor ($\boldsymbol{\tau}$) and heat flux ($\boldsymbol{q}$) are written as,

$$\boldsymbol{\tau} = \mu[\nabla\boldsymbol{u} + (\nabla\boldsymbol{u})^T] - \frac{2}{3}\mu(\nabla \cdot \boldsymbol{u})\boldsymbol{I} \tag{2a}$$

$$\boldsymbol{q} = -\lambda\nabla T - \rho \sum_{k=1}^{N_s} h_k D_k \nabla Y_k \tag{2b}$$

where $T$ is the temperature, $\mu$ is the dynamic viscosity, $\lambda$ is the thermal conductivity, and $h_k$ is the partial enthalpy of species $k$. The total energy is related to the internal energy and kinetic energy, $\rho e_t = \rho e + \frac{1}{2}\rho\boldsymbol{u} \cdot \boldsymbol{u}$. The system is closed with a state equation, $\bar{p} = f(\bar{\rho}, \tilde{T}, \tilde{\boldsymbol{Y}})$, where $T$ is the temperature and sub-grid scale terms are neglected in the evaluation of the pressure. A Vreman sub-grid scale mode is used for the turbulence closure. The sub-grid scale turbulence-chemistry interaction is accounted for by using the dynamic thickened-flame model. The maximum thickening factor is set to be 4 in this study.

### 2.2 Thermodynamic and Transport properties

For computational efficiency and for accurately representing properties near the critical point [12], the Peng-Robinson cubic state equation [13] is used in this study, taking the following form,

$$p = \frac{RT}{v-b} - \frac{a}{v^2 + 2bv - b^2} \tag{3}$$

where $R$ is the gas constant, $v$ is the specific volume, and the coefficients $a$ and $b$ are dependent on temperature and composition to account for the effects of intermolecular forces. Mixing rules are applied for coefficients $a$ and $b$ as follows,

$$a = \sum_{i=1}^{N_s} \cdot \sum_{i=1}^{N_s} X_i X_j a_{ij} \tag{4a}$$

$$b = \sum_{i=1}^{N_s} X_i b_i \tag{4b}$$

where $X_i$ is the mole fraction of species $i$ . The coefficients $a_{ij}$ and $b_i$ are evaluated using the recommended mixing rules from [14],

$$a_{ij} = 0.457236 \frac{(RT_{c,ij})^2}{p_{c,ij}} \left[1 + c_{ij}\left(1 - \sqrt{\frac{T}{T_{c,ij}}}\right)\right]^2 \tag{5a}$$

$$b_i = 0.077796 \frac{RT_{c,i}}{p_{c,ij}} \tag{5b}$$

$$c_{ij} = 0.37464 + 1.54226\omega_{ij} - 0.26992\omega_{ij}^2 \tag{5c}$$

The critical mixture temperature, pressure, molar volume, compressibility, and the acentric factor are determined according to Harstad et al. [14].

Procedures for evaluating thermodynamic quantities such as internal energy, specific heat capacity and partial enthalpy using the Peng-Robinson state equation are described in detail in Ma et al. [11]. The dynamic viscosity and thermal conductivity are evaluated using Chung's method with high-pressure correction [15], and Takahashi's high-pressure correction [16] is used to evaluate binary diffusion coefficients.

### 2.3 Finite-rate chemistry and Chemical Kinetic Model

Chemical source terms in the species conservation equations are defined to be the net production rate from all the reactions which in the present work is modeled by finite-rate chemistry. For large-scale turbulent simulations, the reaction chemistry requires dimensional reduction such that the number of transported scalars is maintained at a reasonable level. In this work, a 33-species reduced mechanism for n-dodecane is used. This mechanism is reduced from a 54-species skeletal mechanism [17] by setting 21 species in QSS approximation. Zero-dimensional auto-ignition computations are performed at a pressure of $p$ = 60 bar, initial temperature in the range of $T$ = 800-1000 K, and equivalence ratio in the range of $\phi$ = 0.5-2. The results are sampled to apply the level of importance criterion [18] using the YARC reduction tool [19]. From this, 21 species are identified to be suitable for quasi-steady-state (QSS) approximation. Linearized quasi-steady state approximation (L-QSSA) is applied to this selection of species. The concentrations of the quasi-steady state (QSS) species are determined by a sparse linear system. A validation is provided in **Fig. 2,** showing that the prediction of ignition time is very close to the original skeletal mechanism for the entire range of temperature and equivalence ratio. The reduced mechanism is incorporated into the CFD solver using a combustion chemistry library based on Cantera. For further details of implementation, please refer to [20].

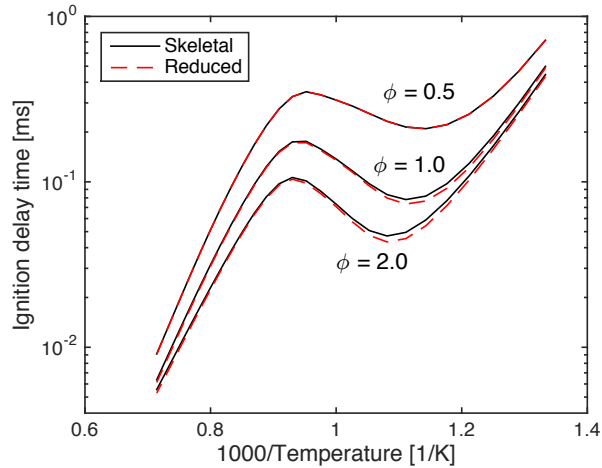

**Fig. 2.** Ignition delay time (ms) of n-dodecane/air mixture predicted by 54-species skeletal mechanism (black solid line) and 33-species reduced mechanism (red dashed line) for three equivalence ratios at $\boldsymbol{p}$ = 60 bar.

*2.4 Numerical methods*

An unstructured finite-volume solver, **CharLESx,** developed at the Stanford Center for Turbulence Research (CTR), is employed in this study [21]. In this solver, the convective fluxes are discretized using a sensor-based hybrid scheme with the entropy-stable flux correction technique developed in Ma et al. [11]. A central scheme, which is 4th-order accurate on uniform meshes, is used along with a 2nd-order ENO scheme. A density sensor is adopted. Due to strong non-linearities inherent in the real-fluid state equation, spurious pressure oscillations will be generated when a fully conservative scheme is used. To eliminate the spurious pressure oscillations, an adaptive double-flux method [9] is employed in this study. A second-order Strang-splitting scheme [22] is applied to separate the convection, diffusion, and reaction operators. A strong stability preserving 3rd-order Runge-Kutta (SSP-RK3) scheme [23] is used for time integration of non-stiff operators. The reaction chemistry is integrated using a semi-implicit Rosenbrock-Krylov (ROK4E) scheme [24], which is 4th-order accurate in time and has linear cost with respect to the number of species. The stability of the ROK4E scheme is achieved through the approximation of the Jacobian matrix by its low-rank Krylov-subspace projection. As few as three right-hand-side evaluations being performed over four stages. Details about the development of the ROK4E scheme in the **CharLES$^x$** solver can be found in Wu et al. [25].

## III.       Results and Discussion

*3.1 Transcritical fuel injection*

In this study, a case denoted "Spray A" is considered, representing a benchmark target from the Department of Energy, Engine Combustion Network, at Sandia National Laboratory. The single hole diesel injection is operated with pure n-dodecane at a rail pressure of 1500 bar. Liquid n-dodecane fuel is injected at 363 K through a nozzle with a diameter of 0.09 mm into a 900 K ambient environment at a pressure of 60 bar. The non-reacting case is first considered with the ambient gas consisting of pure nitrogen. At these conditions, the liquid n-dodecane undergoes a transcritical injection, where the liquid fuel is heated and mixes with the ambient gaseous environment.

A cylindrical computational domain is used in this study with a diameter of 20 mm and a length of 80 mm. In the current study the injector geometry is not included in the computation domain. The mesh is clustered in the region near the injector along the shear layers, and stretched in downstream and radial directions. The minimum grid spacing is 4 μ m, using approximately 20 grid points across the injector nozzle. Fuel mass flux and temperature are prescribed at the injector nozzle using the time-dependent rate of injection modeled using the CMT virtual injection rate generator [12], with default input parameters for the Spray A case (150 MPa, 6 MPa, 0.0894 mm, 0.90, 713.13 kg/m³, and 1.50 ms for injection pressure, back pressure, outlet diameter, discharge coefficient, fuel density and injection time, respectively). Plug flow velocity profile is applied at the nozzle exit without synthetic turbulence. The pressure is prescribed at 6 MPa at the outlet. Adiabatic boundary conditions are applied at all walls. The numerical simulation is initialized with pure nitrogen at 900 K with zero velocity and a pressure of 6 MPa. A CFL number of 1.0 is used during the simulation and a typical time step is about 0.6 ns. The injection process is simulated over a duration of 1.2 ms.

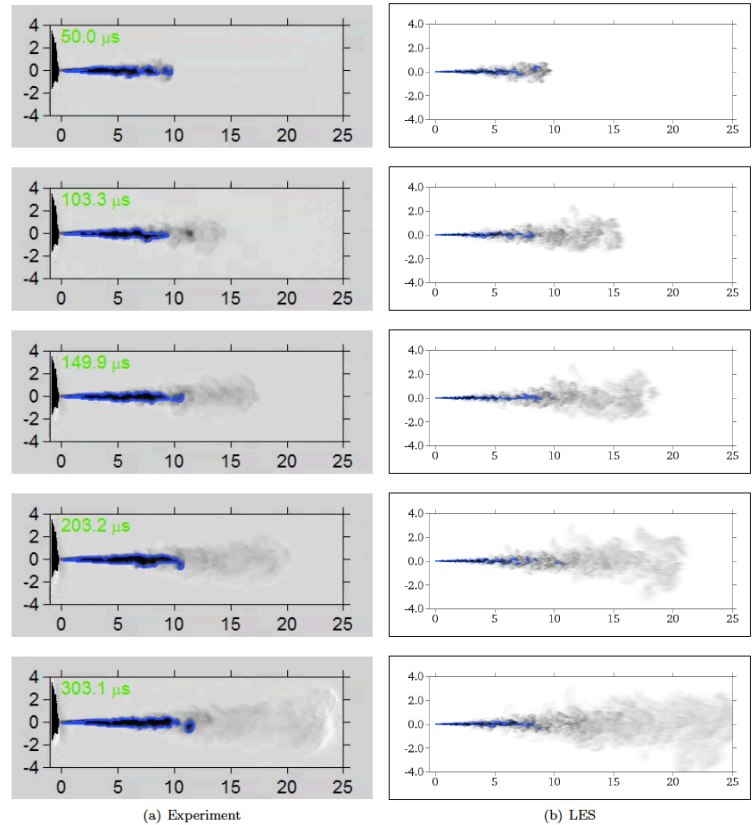

(a) Experiment                (b) LES

**Figure 3.** Comparison between (a) experiment and (b) LES for the injection sequence. Units for spatial dimensions are in mm.

**Figure 3** shows a temporal behavior of the transcritical injection process from the measurements and current simulation results. Experimental images are obtained using diffused back-illumination method with grayscale intensity threshold to indicate liquid region [26,27]. Fuel mass fraction contours at the center-plane are shown in LES results with dense region indicated by mass fraction with a value of 0.6. The simulation results are qualitatively in good agreement with the experimental images in terms of fuel penetration and overall spreading rate. The dense liquid regions are indicated by the blue curves for both the experiment and LES. However, as can be seen from **Fig. 3**, there are some differences in the near nozzle jet fuel spreading behavior. The experimental results show a wider spreading angle near the injector, which may be due to the injection nozzle flow effects that are current neglected in the simulations. Another reason could be due to the application of the diffused interface method near the injector region, where interfaces or droplets, may exist in a transitory state.

The liquid and vapor penetration lengths are extracted from the simulation results using a threshold value of 0.6 and 0.01 for the fuel mass fraction, respectively. The results up to 1 ms after injection are shown in **Fig. 4.** The experimental vapor penetration length determined from Schlieren imaging and liquid penetration length from Mie scattering [28,29], are also shown for comparison. It can be seen that for the vapor penetration an excellent agreement with measurements is obtained. It was found that the utilization of the inflow boundary condition with a time-dependent fuel mass flux is essential for the accurate prediction. A sensitivity study on the numerical threshold values of the liquid penetration, where a value of 0.6 is used for the mixture fraction, revealed it to be somewhat arbitrary. It was found that for values from 0.95 to 0.4 the predicted length varies between 5 mm to 15 mm. This is similar to what was reported in measurements, where the solution depends of the choice of measurement method, the optical setup, the criteria of liquid determination, and the geometry of the injector. The resolution in either technique is still not close to fully resolve the smallest flow structures near the injector.

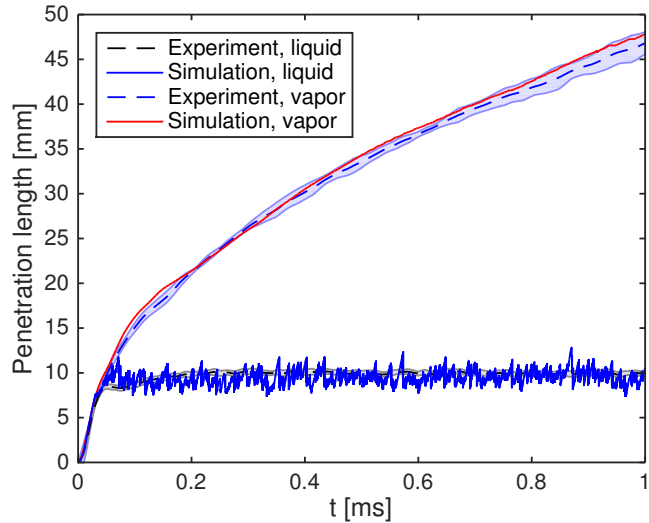

**Figure 4**. Liquid and vapor penetration lengths predicted in comparison with experimental data [13].

The flow structures and mixing behaviors of the injection process further downstream are compared to the measurements of mixture fraction by Rayleigh scattering. Multiple injections in the experiments provide ensemble-averaged statistics. In the simulation, the statistics of the steady period of injection are obtained by temporally averaging between 0.6 ms and 1.2 ms after the injection. Figure 14 shows a comparison of the radial mixture fraction distribution at two different axial locations (x = 25, and 35 mm). As can be seen in **Fig. 5**, there is a good agreement in the mean values of the mixture fraction at all three locations, while the simulation predicts slightly higher rms values compared to the experimental data. These results along with the excellent agreement of the vapor penetration length as presented, show that the current numerical method is capable of predicting the turbulent mixing process between fuel and surrounding environment downstream of the injector after the dense liquid fuel is fully disintegrated.

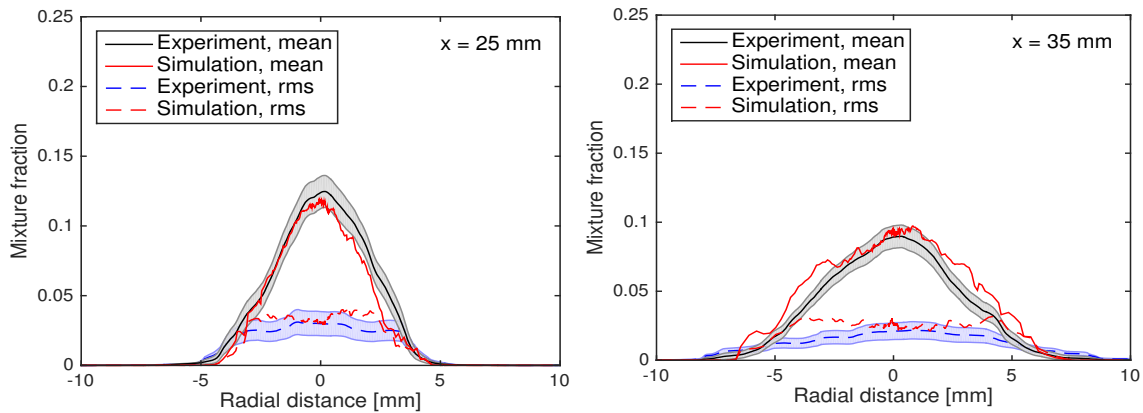

**Figure 5.** Radial profiles of mean and rms values of mixture fraction at two different axial locations in comparison with the experimental data measured by Rayleigh scattering [14].

The non-reacting spray simulations presented thus far establish the fidelity of the solver showing the importance of heat transfer and turbulent mixing. The remainder of the paper will focus on the auto-ignition characteristics of n-dodecane fuel at various operating points in the low and high temperature combustion regimes in the range of $750K < T < 1200K$ with experimental comparison in terms of ignition delay and lift-off.

### 3.2. Transcritical auto-ignition

Following the evaluation of the solver for the non-reacting case, we now turn our attention to reacting cases. The simulation results for the case with an ambient temperature of 900 K (Spray A conditions) are shown in Fig. 6. Temperature fields are presented at several injection times along with the $CH_2O$ and OH fields. The results for $CH_2O$ and OH fields are plotted in same figures to emphasize the spatial separation between them. It can be seen from the temperature results that the liquid *n*-dodecane fuel jet is heated up by the surrounding hot environment after being injected into the combustion chamber and a first-stage ignition can be observed at early ignition times (e.g., see temperature fields at 300 μs), which is associated with the rise in temperature and the formation of the $CH_2O$ species. After the second-stage ignition process (see results in Fig. 6 after 300 μs), high temperature regions with temperatures over 2000 K can be seen downstream the combustion chamber. From the species results in **Fig. 6**, it can be seen that $CH_2O$ is formed initially at the radial periphery of the jet. At later times, the maximum concentration of $CH_2O$ is observed in the center of the penetrating jet. The formation of OH is associated with the subsequent consumption of $CH_2O$ [30] and the high-temperature chemistry (HTC) corresponding to the main second-stage ignition process. High concentrations of OH are found near the edges of the penetrating jet due to the relatively low scalar dissipation rate and longer residence time in these regions [8].

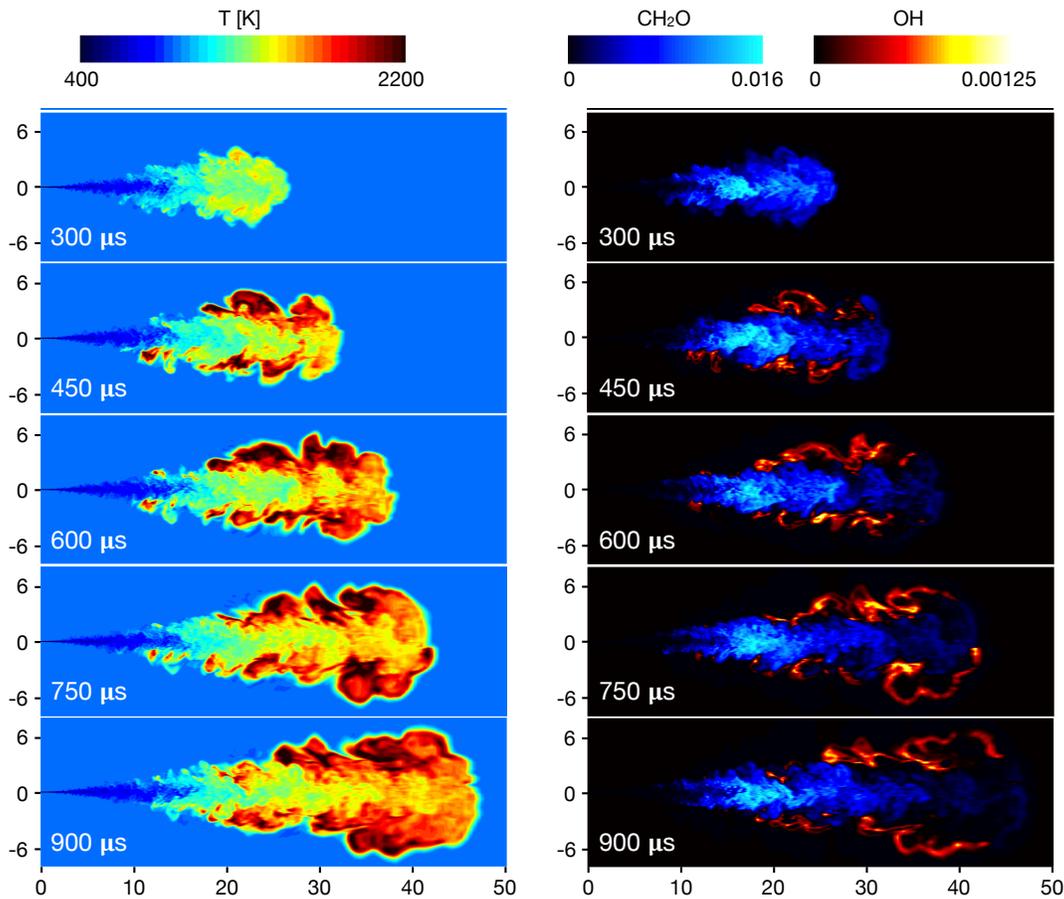

**Figure 6.** Auto-ignition sequence of the case with 900 K ambient temperature (Spray A conditions) showing temperature and intermediate species fields. Spatial units in mm

To assess the model performance in predicting intermediate species during the auto-ignition processes, Figs 7 and 8 show results for mass fractions of $CH_2O$ and $OH$ species along the center-plane at several injection times for the 900 K case in comparison with PLIF measurements [31]. It can be seen from Fig. 7 that there is a good agreement between LES and measurements in terms of the shape, magnitude, and location of the formation of $CH_2O$. A good agreement in OH fields can be observed between LES and experiments in Fig. 8, with a sharper representation in the simulation results.

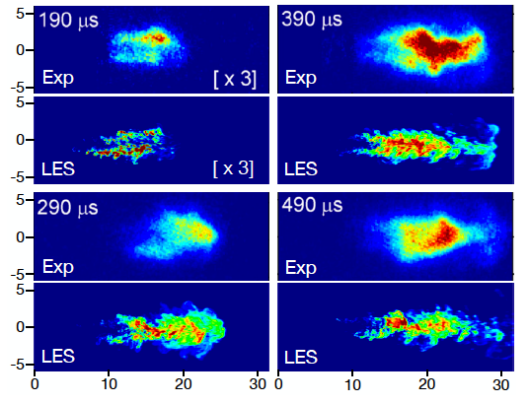

**Figure 7.** Comparison between $CH_2O$ mass fraction from LES results and false-color PLIF measurements at several injection times at 900 K ambient temperature. Spatial units in mm.

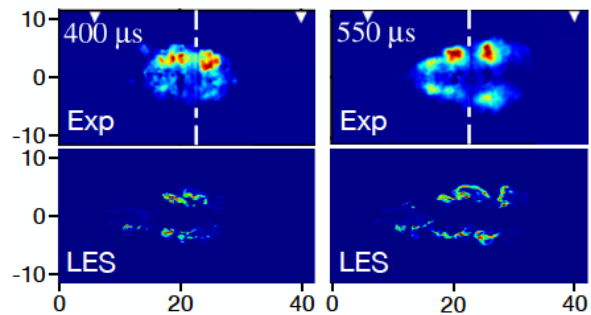

**Figure 8.** Comparison of OH mass fraction from LES results and OH PLIF measurements at two injection times at 900 K ambient temperature. Spatial units in mm.

**Figure 9** shows the ignition delay time at different ambient temperature conditions, predicted by LES compared with measurements [28, 33]. Following the ECN-recommendations and criteria used by previous studies [6, 8, 32], the ignition delay time in LES is defined as the time when the maximum OH mass fraction reaches 14% of the value at quasi-steady state of the flame. As can be seen from Fig. 9, good agreement is observed between LES and experiments, with about 10% maximum error from LES. Shorter ignition delay times were also predicted by Yao et al. [17] where the same parent skeletal chemical mechanism was adopted. Previous work utilizing flamelet-based combustion models [6] also showed that the chemical mechanism has a significant effect on accurately predicting the ignition delay time. Note, that although there is a larger discrepancy with the simulations at the lowest operating point (800K), the experimental uncertainty is significant at this stage $ID_{800K}$ = 1.2 ms +/- 0.3 ms, as reported in [33].

**Figure 10** below, shows a cut-plane 3D rendering of the OH and $CH_2O$ fields and their interaction in the main stage ignition process at a chamber temperature of $800K$. It shows the spatial separation between both fields with growth of OH surrounding the CH2O fields in the center of the spray. The fields and their intensities are shown between 0.825 and 1.05 ms.

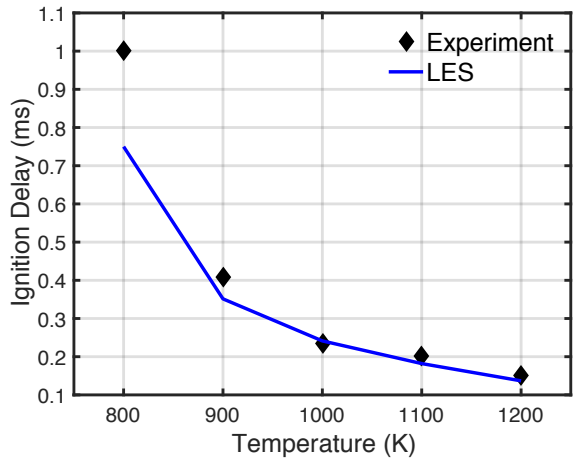

**Fig. 9.** Ignition delay time predicted by LES at different ambient temperature conditions in comparison with measurements.

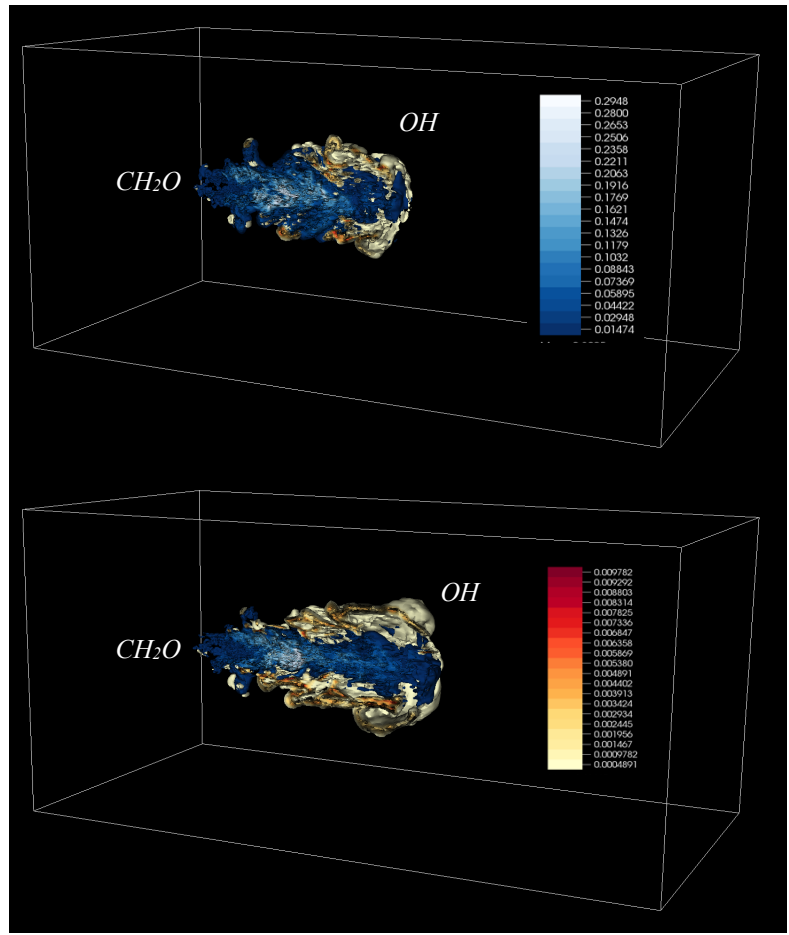

**Fig. 10.** Cut-plane contour rendering of the interaction between formaldehyde (CH20) and hydroxide (OH) fields at (top) 0.825 ms, and (bottom) 1.05 ms at 800K chamber temperature.

## IV.   Conclusions

A diffuse-interface method in conjunction with a finite-rate chemistry LES model is presented for the modeling of diesel fuel injection and auto-ignition processes under transcritical injection conditions. Compressible multi-species conservation equations are solved with a Peng-Robinson state equation and real-fluid transport properties. LES-calculations are performed to simulate the ECN Spray A benchmark configuration for both inert and reacting conditions. The simulations considered target cases of *n*-dodecane spray non-reacting mixing and reacting conditions from 800*K* to 1200*K* and compared with available experimental measurements. For the non-reacting case, predicted vapor and liquid penetration lengths are in good agreement with experimental results. Five ambient temperature conditions are considered for reacting cases. Good agreement of the ignition delay time is obtained from simulation results at different ambient temperature conditions and the formation of intermediate species is captured by the simulations, indicating that the presented numerical framework adequately reproduces the corresponding LTC and HTC ignition processes under high-pressure conditions that are relevant to realistic diesel fuel injection.

## ACKNOWLEDGMENTS

This research is supported in part by resources from DoD High Performance Computing Modernization Program (HPCMP) FRONTIER Award at ARL through project "Petascale High Fidelity Simulation of Atomization and Spray/Wall Interactions at Diesel Engine Conditions" and accomplished under Cooperative Agreement #W911NF-16-2-0170.